\documentclass[aps,amssymb,pra,showpacs,twocolumn]{revtex4}
\usepackage{epsf}
\usepackage{amsmath}
\usepackage{graphicx}
                                                                                
\begin{document}                                                       
                                                                 
\title{Underbarrier interference}
                          
\author{B. Ivlev} 

\affiliation
{Instituto de F\'{\i}sica, Universidad Aut\'onoma de San Luis Potos\'{\i}, San Luis Potos\'{\i}, Mexico,\\
Department of Physics and Astronomy and NanoCenter, University of South Carolina, Columbia, South Carolina, USA}

%\date{\today}

\begin{abstract}
Quantum tunneling through a two-dimensional static barrier becomes unusual when a momentum of an electron has a tangent component with 
respect to a border of the prebarrier region. If the barrier is not homogeneous in the direction perpendicular to tunneling a fraction of 
the electron state is waves propagating away from the barrier. When the tangent momentum is zero a mutual interference of the waves results 
in an exponentially small outgoing flux. The finite tangent momentum destroys the interference due to formation of caustics by the waves. As
a result, a significant fraction of the prebarrier density is carried away from the barrier providing a not exponentially small penetration 
even through an almost classical barrier. The total electron energy is well below the barrier. 
    
\end{abstract} \vskip 1.0cm
   
 
\maketitle
\section{Introduction}
The problem of tunneling through a potential barrier was a subject of attention for many years in nuclear physics, particle physics, atomic
physics, and chemical physics. Quantum tunneling across a one-dimensional static potential barrier is described by the theory of Wentzel, 
Kramers, and Brillouin (WKB) \cite{LANDAU} if the barrier is not very transparent. Early studies of quantum tunneling were performed in
Refs.~\cite{GAM,CON,LAU}. Some modern aspects of the phenomenon can be found, for example, in Refs.~\cite{PER,KAG,COMPL,ANK}. 

\begin{figure}
\includegraphics[width=4.5cm]{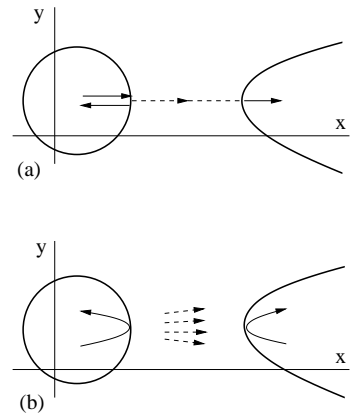}
\caption{\label{fig1}Solid curves correspond to a constant potential energy. Tunneling occurs between two classically allowed regions, the
prebarrier one is to the left. (a) The conventional mechanism. The main trajectory in the classically forbidden region is indicated by the 
dashed path. (b) The electron hits the border with a tangent velocity. The set of subsequent propagating waves is shown by the dashed 
arrows.} 
\end{figure}

A scenario of tunneling through a multi-dimensional barrier is well studied 
\cite{LIFSHITZ,POKR,VOL,STONE,COLEMAN,COLEMAN1,COLEMAN2,MILLER,SCHMID1,SCHMID2,DYK,GURV}. The main contribution to a tunneling probability
comes from the extreme path in the $\{x,y\}$ plane (we consider two dimensions) linking two classically allowed regions as drawn in 
Fig.~\ref{fig1}(a). The path is a classical trajectory with real coordinates which can be parameterized by imaginary time. The underbarrier 
trajectory is a solution of Newton's equation in imaginary time. The trajectory is given rise by a particle hitting with a zero tangent 
momentum a border of the classically allowed region from the the classical side. This is shown in Fig.~\ref{fig1}(a). Under the barrier 
the probability density reaches a maximal value at each point of the trajectory along the orthogonal direction with respect to it. 
Therefore around the trajectory, which plays a role of a saddle point, quantum fluctuations are weak. The wave function, tracked along 
that trajectory under the barrier, exhibits an exponential decay generic with WKB behavior. 

However, in some cases tunneling through multi-dimensional barriers occurs according to a different scenario which is far from being similar
to WKB. In this paper unusual tunneling through two-dimensional barriers is studied.  

When the prebarrier state has a tangent component of a momentum, as in Fig.~\ref{fig1}(b), there are no extreme points at the border of 
the prebarrier region since the derivative of a wave function along the border is finite. This means that a tunneling probability is no more
determined by the main underbarrier path but comes from a wide set of paths indicated in Fig.~\ref{fig1}(b) by the dashed arrows. 
Traditionally, a decay of a state with a tangent momentum is not considered since it does not correspond to a saddle point and, hence, the 
net contribution is supposed to be averaged down to a small value. As shown in the paper, that conclusion is not correct and states with 
tangent momentum can play a crucial role in tunneling processes. 

The phenomenon can be explained in terms of the following arguments.

When a two-dimensional barrier is homogeneous perpendicular to the tunneling direction $x$ ($V_{0}(x)$) tunneling is generic with WKB 
mechanism. Suppose that there is an impurity localized at the barrier region and described by the potential $u(x,y)$. The underbarrier wave 
function in the total potential $V_{0}(x)+u(x,y)$ contains a set of overbarrier propagating waves if to treat them as eigenfunctions of 
$V_{0}(x)$. When a tangent momentum is zero, as in Fig.~\ref{fig1}(a), the energy distribution of the propagating waves is smooth leading to
a strong mutual cancellation of their contributions to a total wave function due to interference. As a result, a contribution of the 
propagating waves to an outgoing flux is exponentially small. 

In contrast, when a tangent momentum is finite, as in Fig.~\ref{fig1}(b), the mutual cancellation of those waves due to interference can be 
not complete. Analogous to optics, de Broglie waves may form caustics where a distribution of waves becomes not smooth
\cite{LANDAU1,SCHMID1,SCHMID2,DYK}. This violates the strong mutual cancellation of the propagating waves and the surviving fraction of
them goes away from the barrier providing a not small output. More details are given in Sec.~\ref{nature}. 

The underbarrier interference opens a possibility of penetration through almost classical barriers. The enhanced tunneling through a static 
barrier was initially proposed  in Ref.~\cite{IVLEV6}. This phenomenon was also studied in tunneling through nonstationary barriers 
\cite{IVLEV2,IVLEV3,IVLEV4,IVLEV5,IVLEV8} where an underbarrier phase was created by quanta emission. 

An exact wave function in tunneling problem can be written in the form $C(x,y)\exp[iS(x,y)/\hbar]$ where $S$ is a classical action and $C$ 
is a prefactor. An exact analytical solution of the Schr\"{o}dinger equation for a complicated barrier is impossible. We use a semiclassical
approach when only the exponential part of the wave function is calculated. This is called the main exponential approximation when the 
action $S$ is large and the prefactor $C$ is less important. The classical action is calculated in Sec.~\ref{sol} as a solution of the 
Hamilton-Jacobi equation. The action was also tracked along classical trajectories in imaginary time. The both methods lead to the same 
tunneling probability. 
           
At the first sight, the obtained solution of the Hamilton-Jacobi equation provides the main (exponential) part of the exact wave function 
with no problems. But there is a delicate phenomenon in quantum mechanics which may destroy a semiclassical solution. When two branches of a
wave function exist at the same spatial domain the smaller one can unexpectedly disappear. From the semiclassical standpoint it happens as a
jump in space of the prefactor $C(x,y)$ down to zero. This is called Stokes phenomenon \cite{STOKES,HEAD,MASL}. In Sec.~\ref{2D} it is shown
that there are no Stokes jumps in the semiclassical solution obtained. So despite the exact solution is unknown, it is proved that 
$\exp[iS(x,y)/\hbar]$ is the main part of the exact wave function and the prefactor plays a secondary role.

In principle, a numerical solution of the static Schr\"{o}dinger equation in two dimensions is possible. But in our case one should keep at 
some spatial domains two branches when one of them is exponentially small. In order to resolve that branch a precision of calculations has 
to be very high resulting in an extremely large number of discrete points in a two-dimensional net using for a numerical computation. This 
essentially exceeds memory facilities of computers. So a numerical study of the problem seems to be impossible at present. 

In Sec.~\ref{imp} tunneling from a quantum wire through a barrier with an impurity is investigated. An influence of impurities inside a 
potential barrier on tunneling was widely studied. See, for example, \cite{LIF,SHKL1,MESH,SHKL2}. A famous mechanism is resonant (Wigner) 
tunneling when an impurity level coincides with a particle energy \cite{LANDAU}. This is not our case since we are away of Wigner resonance.
Another famous mechanism is the interference in scattering in a system of many impurities as in localization phenomena \cite{GANT}. This is 
also not our case since there is just one underbarrier scattering center. 

The proposed mechanism of impurity assisted tunneling was in shadow in the previous studies. A necessary condition is a tangent momentum in 
a prebarrier region. Underbarrier waves with the tangent momentum are scattered by the impurity resulting in overbarrier propagating waves 
which carry away a significant fraction of the prebarrier density.
\section{FORMULATION OF THE PROBLEM}
\label{form}
Below we consider a simpler barrier than one in Fig.~\ref{fig1}. Tunneling occurs in the $x$ direction from a straight infinitely long 
quantum wire aligned along the $y$ axis. A particle, localized in the $\{x,y\}$ plane, is described by the Schr\"{o}dinger equation 
\begin{eqnarray} 
\nonumber
-\frac{\hbar^{2}}{2m}\left(\frac{\partial^{2}\psi}{\partial x^{2}}+\frac{\partial^{2}\psi}{\partial y^{2}}\right)
-\hbar\sqrt{\frac{2u_{0}}{m}}\sqrt{1+\beta^{2}(y)}\,\delta(x)\psi\\
\label{1}
-{\cal E}|x|\psi=E\psi.
\end{eqnarray}
The $\delta$ well in Eq.~(\ref{1}) relates to the quantum wire placed at the position $x=0$. When the wire is homogeneous ($\beta(y)=0$) the
discrete energy level in the $\delta$ well is $-u_{0}$. The non-homogeneity of the wire is supposed to be localized at a finite region of
$y$ where $\beta(y)$ is not zero and therefore $\beta(\pm\infty)=0$. We also suppose that $\beta(y)=\beta(-y)$. Tunneling occurs through
the triangular potential barrier created by the static electric field ${\cal E}$. The energy is below the barrier ($E<0$) and in order to 
move from the wire ($x=0$) to an infinite $x$ one should pass through the potential barrier. For simplicity we use the even potential 
$-{\cal E}|x|$ to get the problem symmetric with respect to $x$. One can consider positive $x$ only if to put the boundary 
condition 
\begin{equation} 
\label{1a}
\frac{\partial\psi(x,y)}{\partial x}\bigg|_{x=0}=-\frac{\sqrt{2mu_{0}}}{\hbar}\sqrt{1+\beta^{2}(y)}\,\psi(0,y),
\end{equation}
which follows from Eq.~(\ref{1}). 
\section{ZERO ELECTRIC FIELD}
\label{zero}
First, let us find a wave function for a wide barrier when the electric field ${\cal E}$ is zero. In this case the barrier becomes 
rectangular. By Fourier components the wave function reads
\begin{eqnarray} 
\nonumber
&&\psi(x,y)=\int\psi(0,y_{1})dy_{1}\\
\label{1b}
&&\int\frac{dk}{2\pi}\exp\Bigg[ik(y-y_{1})-x\sqrt{k^{2}-\frac{2mE}{\hbar^{2}}}\Bigg].
\end{eqnarray}
An equation for $\psi(0,y)$ can be obtained in the same manner through Fourier components from Eq.~(\ref{1a})
\begin{eqnarray} 
\label{112a}
\int dy_{1}\psi(0,y_{1})\int\frac{dk}{2\pi}e^{ik(y-y_{1})}\sqrt{\frac{\hbar^{2}k^{2}}{2mu_{0}}-\frac{E}{u_{0}}}\\
\nonumber
=\sqrt{1+\beta^{2}(y)}\psi(0,y).
\end{eqnarray}
Eqs.~(\ref{1b}) and (\ref{112a}) provide an exact description of exponential decay under the rectangular barrier.

In the semiclassical limit the solution of Eq.~(\ref{112a})
\begin{equation} 
\label{112b}
\psi(0,y)\sim\exp\left[i\frac{\sqrt{2mu_{0}}}{\hbar}\int^{y}_{0}dy_{1}\sqrt{1+\frac{E}{u_{0}}+\beta^{2}(y_{1})}\right]
\end{equation}
is obtained with exponential accuracy by a saddle method. Validity of the solution (\ref{112b}) can be easily checked if to insert it into 
Eq.~(\ref{112a}) and to take saddles subsequently with respect to $y_{1}$ and $k$. 

The physical meaning of a state in the non-homogeneous wire when an electric field is zero becomes clear in the limit when $\beta(y)\ll 1$
and $(E+u_{0})\ll u_{0}$. In this case the wave vector $k$ is small and Eq.~(\ref{1b}) reads
\begin{equation} 
\label{112c}
\psi(x,y)=\psi(0,y)\exp\left(-\frac{x}{\hbar}\sqrt{2mu_{0}}\right).
\end{equation}
Eq.~(\ref{112a}) becomes of a Schr\"{o}dinger form
\begin{equation} 
\label{112d}
-\frac{\hbar^{2}}{2m}\frac{\partial^{2}\psi(0,y)}{\partial y^{2}}-u_{0}\beta^{2}(y)\psi(0,y)=(E+u_{0})\psi(0,y).
\end{equation}
When $E>-u_{0}$ a solution of Eq.~(\ref{112d}) at a large negative $y$ is a plane wave which reflects from the effective potential 
$-u_{0}\beta^{2}(y)$. This case is generic with an overbarrier reflection \cite{LANDAU}. At $|y|\rightarrow\infty$
\begin{eqnarray} 
\nonumber
&&\psi(0,y)=\exp\left[\frac{iy}{\hbar}\sqrt{2m(E+u_{0})}\right]\\
\label{112e}
&&+\theta(-y)\sqrt{R}\exp\left[-\frac{iy}{\hbar}\sqrt{2m(E+u_{0})}\right].
\end{eqnarray}
When the typical distance $y\sim a$, where $\beta(y)$ is not zero, is sufficiently large the reflection coefficient $|R|$ in 
Eq.~(\ref{112e}) is exponentially small \cite{LANDAU}
\begin{equation} 
\label{112f}
|R|\sim\exp\left[-c\,\frac{a}{\hbar}\sqrt{2m(E+u_{0})}\right],\hspace{0.4cm}0<c\sim 1.
\end{equation}
\section{FINITE ELECTRIC FIELD} 
\label{finite}
Below we use the dimensionless form of the Schr\"{o}dinger equation (\ref{1}) measuring $x$ and $y$ in the units of $u_{0}/{\cal E}$
\begin{eqnarray} 
\label{2}
-\frac{1}{B^{2}}\left(\frac{\partial^{2}\psi}{\partial x^{2}}+\frac{\partial^{2}\psi}{\partial y^{2}}\right)
-\frac{2}{B}\sqrt{1+\alpha^{2}(y)}\,\delta(x)\psi\\
\nonumber
-|x|\psi=(\gamma-1)\psi.
\end{eqnarray}
The large semiclassical parameter is
\begin{equation} 
\label{3}
B=\frac{u_{0}\sqrt{2mu_{0}}}{\hbar{\cal E}},
\end{equation}
$\alpha(y)=\beta({yu_{0}/\cal E})$, and the energy is $E=(\gamma-1)u_{0}$. Since the energy $E$ is below the barrier the condition 
$\gamma<1$ holds. The equation (\ref{2}) can be considered at $x>0$ if to impose the boundary condition (\ref{1a}) which in the 
dimensionless units reads
\begin{equation} 
\label{3a}
\frac{\partial\psi(x,y)}{\partial x}\bigg|_{x=0}=-B\sqrt{1+\alpha^{2}(y)}\,\psi(0,y).
\end{equation}
At positive $x$ the variables in Eq.~(\ref{2}) are separated and a solution can be written in the general form
\begin{eqnarray} 
\label{15}
&&\psi(x,y)=\int_{C}dk\exp[iBF(k)]\\
\nonumber
&&\exp\left(iBy\sqrt{\gamma+k^{2}}-B\int^{x}_{0}dx_{1}\sqrt{1+k^{2}-x_{1}}\right).
\end{eqnarray}
The integration contour(s) $C$ lies in the plane of complex variable $k$. This relates to Laplace's method for differential equations.
The function $F(k)$ should be determined from the condition (\ref{3a}). The resulting equation for 
$\psi(0,y)=\int dk\exp(iBF+iBy\sqrt{\gamma+k^{2}})$ is generic with Eq.~(\ref{112a}). 

A calculated function $F(k)$ has to be inserted into Eq.~(\ref{15}). Since the parameter $B$ is large one can use a saddle point method to
evaluate the $k$ integral in Eq.~(\ref{15}). For each $x$ and $y$ this method gives a certain saddle value $k_{s}(x,y)$. Quantum 
fluctuations around the saddle determine a preexponential factor. As shown below, the saddle formalism corresponds to a general integral of 
Hamilton-Jacobi equation considered in Sec.~\ref{ham-jac}. 
\section{HAMILTON-JACOBI FORMALISM}
\label{ham-jac}
In this section we develop a semiclassical method of solution of Eq.~(\ref{2}) which is reduced to the Hamilton-Jacobi formalism. The wave 
function is of the form 
\begin{equation} 
\label{4}
\psi(x,y)=\exp\left[iB\sigma(x,y)\right].
\end{equation} 
If to insert the expression (\ref{4}) into Eq.~(\ref{2}) we obtain at positive $x$ the equation 
\begin{equation} 
\label{52}
\left(\frac{\partial\sigma}{\partial x}\right)^{2}+\left(\frac{\partial\sigma}{\partial y}\right)^{2}-x
-\frac{i}{B}\left(\frac{\partial^{2}\sigma}{\partial x^{2}}+\frac{\partial^{2}\sigma}{\partial y^{2}}\right)=\gamma-1.
\end{equation}
Since $B$ is a large parameter Eq.~(\ref{52}) is reduced to 
\begin{equation} 
\label{5}
\left(\frac{\partial\sigma}{\partial x}\right)^{2}+\left(\frac{\partial\sigma}{\partial y}\right)^{2}-x=\gamma-1.
\end{equation}
Eq.~ (\ref{5}) is called the equation of Hamilton-Jacobi for the classical action $S=\hbar B\sigma(x,y)$. The function $S(x,y)$ is really
classical one since it does not depend on Planck's constant as $B$ is inversely proportional to $\hbar$. Eq.~(\ref{5}) holds at positive 
$x$. The boundary condition follows from Eq.~(\ref{3a})
\begin{equation} 
\label{101}
\frac{\partial\sigma(x,y)}{\partial x}\bigg|_{x=0}=i\sqrt{1+\alpha^{2}(y)}.
\end{equation}
A general integral of the Hamilton-Jacobi equation can be obtained by the method of variation of constants \cite{LANDAU2}. The general 
integral, satisfying the condition (\ref{101}), has the form
\begin{eqnarray} 
\label{7}
&&\sigma(x,y)=i\int^{x}_{0}dx_{1}\sqrt{\alpha^{2}[iv(x,y)]+1-x_{1}}\\
\nonumber
&&+y\sqrt{\gamma+\alpha^{2}[iv(x,y)]}-\int^{iv(x,y)}_{0}y_{1}\frac{\partial\sqrt{\gamma+\alpha^{2}(y_{1})}}{\partial y_{1}}\,dy_{1},
\end{eqnarray}
where the function $v(x,y)$ obeys the equation
\begin{equation} 
\label{8}
\frac{v(x,y)+iy}{\sqrt{\gamma+\alpha^{2}[iv(x,y)]}}=\int^{x}_{0}\frac{dx_{1}}{\sqrt{\alpha^{2}[iv(x,y)]+1-x_{1}}}.
\end{equation}
In this formalism the relations hold 
\begin{eqnarray}
&&\label{9}
\frac{\partial\sigma(x,y)}{\partial x}=i\sqrt{\alpha^{2}[iv(x,y)]+1-x},\\
\label{10}
&&\frac{\partial\sigma(x,y)}{\partial y}=\sqrt{\gamma+\alpha^{2}[iv(x,y)]}.
\end{eqnarray}
As follows from Eq.~(\ref{8}), $iv(0,y)=y$ and therefore $\partial\sigma(0,y)/\partial y=\sqrt{\gamma+\alpha^{2}(y)}$. To obtain the 
function $\sigma(x,y)$ one has to determine the function $v(x,y)$ from Eq.~(\ref{8}) and to insert it into Eq.~(\ref{7}).

Eq.~(\ref{8}) is a condition of independence of $\sigma$ on ``constant'' $v(x,y)$. It is  expressed by the relation 
$\partial\sigma/\partial v=0$ which is generic with the saddle condition $k=k_{s}(x,y)$ in Eq.~(\ref{15}). We see that 
$k_{s}(x,y)=\alpha[iv(x,y)]$ and the last term in Eq.~(\ref{7}) is $-F(k)$ if to express $v$ through $k$ from the equation $k=\alpha(iv)$.
\begin{figure}
\includegraphics[width=4.5cm]{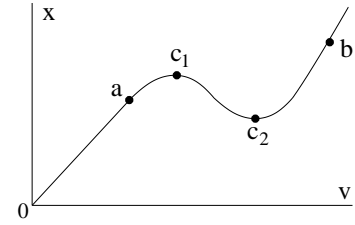}
\caption{\label{fig2}The plot of $x$ versus $v$ at $y=0$ according to Eq.~(\ref{103}) at $a_{0}<a$. See Sec.~\ref{sol}.} 
\end{figure}
\section{SOLUTION}
\label{sol}
After a little algebra the equation (\ref{8}) reads
\begin{equation} 
\label{103}
x=(v+iy)\sqrt{\frac{1+\alpha^{2}(iv)}{\gamma+\alpha^{2}(iv)}}
-\frac{(v+iy)^{2}}{4[\gamma+\alpha^{2}(iv)]}.
\end{equation}
Since $\alpha(y)$ is an even function there is a real solution $v(x,0)$ of Eq.~(\ref{103}). Therefore there is an imaginary branch of
$\sigma(x,0)$ as follows from Eq.~(\ref{7}). 

\begin{figure}
\includegraphics[width=5.5cm]{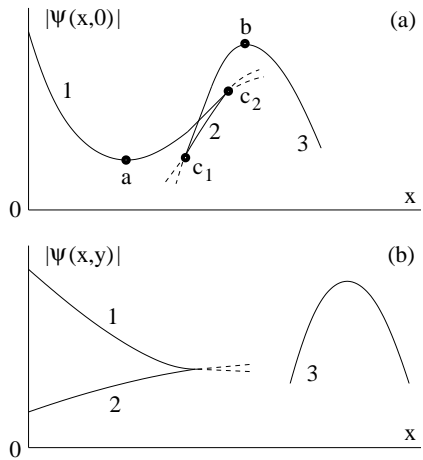}
\caption{\label{fig3}Branches of the wave function at $a_{0}<a$. (a) At $y=0$ there are two branching points $c_{1}$ and $c_{2}$. (b) The
case of a large $y$. The branches 1 and 2 at a small $x$ are generic with conventional WKB ones in a one-dimensional case. The top of the 
branch 3 moves as a classical particle to infinity.} 
\end{figure}

We specify now the particular form 
\begin{equation} 
\label{103a}
\alpha(y)=\alpha_{0}\exp\left(-\frac{y^{2}}{a^{2}}\right).
\end{equation}
One can easily plot $x$ as a function of $v$ at $y=0$. At $\alpha^{2}_{0}=0.03$, $\gamma=0.2$, and $a=2$ the plot is shown in 
Fig.~\ref{fig2}. It determines the real function $v(x,0)$. The curve in Fig.~\ref{fig2} has two special points marked as $c_{1}$ and 
$c_{2}$. The variable $x$ cannot be extended above the point $c_{1}$ and below the point $c_{2}$ in frameworks of a real $v$. The same is 
valid for the action $\sigma$ and, therefore, for the wave function (\ref{4}) plotted in Fig.~\ref{fig3}(a). 

In Fig.~\ref{fig3}(a) $c_{1}$ and $c_{2}$ are branching points where two of three branches 1, 2, and 3 merge. Extensions of the branches are
shown by the dashed curves corresponding to extensions of $v(x,0)$ to the complex plane above $c_{1}$ and below $c_{2}$ in Fig.~\ref{fig2}. 
Near the point $c_{2}$ one can estimate
\begin{eqnarray} 
\label{103aa}
\nonumber
&&i\sigma(x,0)-i\sigma(x_{c_{2}},0)-(x_{c_{2}}-x)\sqrt{\alpha^{2}(iv_{c_{2}})+1-x_{c_{2}}}\\
&&\sim\pm(x_{c_{2}}-x)^{3/2}.
\end{eqnarray}
Analogous relation holds for the point $c_{1}$. 

At the points $a$ and $b$ the function $\psi(x,0)$ reaches extrema. According to Eq.~(\ref{9}), at the point $b$
\begin{equation} 
\label{103b}
x_{b}=1+\alpha^{2}(iv_{b}),
\end{equation}
where $v_{b}$ is determined by the largest root of Eq.~(\ref{103})
\begin{equation} 
\label{103c}
v_{b}=2\sqrt{1+\alpha^{2}(iv_{b})}\sqrt{\gamma+\alpha^{2}(iv_{b})}.
\end{equation}
Equations analogous to (\ref{103b}) and (\ref{103c}) also hold for the point $a$ when one should take the smaller real root of 
Eq.~(\ref{103c}). The points $a$ and $b$ are marked in Fig.~\ref{fig2}.

What happens to $\psi(x,y)$ at a finite $y$? To answer this question we consider a classical trajectory in the potential $(-x)$ relating to
the total energy $\gamma-1$
\begin{equation} 
\label{103d}
x=x_{b}+\frac{y^{2}}{4(x_{b}-1+\gamma)}.
\end{equation}
As follows from the equation (\ref{103}), $v(x,y)=v_{b}$ if $x$ and $y$ are connected by the relation (\ref{103d}). According to 
Eqs.~(\ref{9}) and (\ref{10}), imaginary part of $\sigma(x,y)$ is a constant on the classical trajectory (\ref{103d}). This means that the
extreme point of the branch 3 in Fig.~\ref{fig3} moves in the plane $\{x,y\}$ along the classical trajectory (\ref{103d}) where $|\psi|$
keeps a constant value. This statement is valid in the main exponential approximation used when a quantum mechanical smearing of packets is 
negligible. 

Along the part $y<0$ of the branch 3 the particle flux is directed towards the well and at $0<y$ it has the opposite direction. So the total 
flux of particles, associated with the branch 3, in the $x$ direction is zero but in the $y$ direction it has the same sign as the flux 
along the $\delta$ well. 

In Fig.~\ref{fig3}(b) the branches are shown at a large $y$ when the $\delta$ well is homogeneous in the $y$ direction. In this case the
branches 1 and 2, at a not large $x$, are generic with conventional WKB ones in one dimension. The dashed part of the branches 1 and 2 
corresponds to an exponentially small outgoing flux away from the barrier. The branch 3 in Fig.~\ref{fig3}(b) is located at a large $x$ 
corresponding to a far point of the classical trajectory (\ref{103d}).

\begin{figure}
\includegraphics[width=5.5cm]{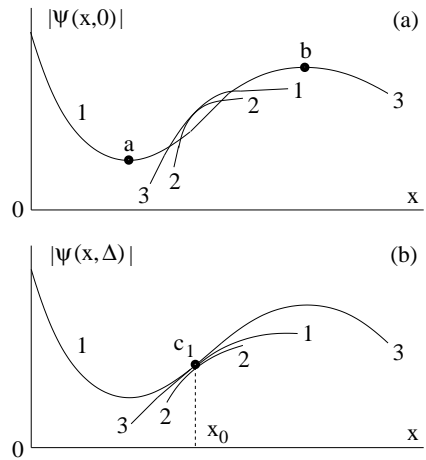}
\caption{\label{fig4}Branches of the wave function at $a<a_{0}$. (a) Hybridized branches 1-3 and 3-1 and the branch 2-2 at $y=0$. (b) 
Branches 1-1, 2-2, and 3-3 touch each other at the singular point $c_{1}$ when $y=\Delta$. Another singular point $c_{2}$ corresponds to 
$y=-\Delta$.} 
\end{figure}

The behavior of branches of the wave function in Fig.~\ref{fig3} occurs when the width $a$ of the $\delta$ well profile is not too small. 
Under reduction of $a$ a branch behavior becomes qualitatively different. Namely, the region between the points $c_{1}$ and $c_{2}$ shrinks 
and becomes zero at $x=x_{0}$ and $v=v_{0}$ when $a=a_{0}$. Note that the points $a$ and $b$ remain well separated. For $\alpha_{0}$ and 
$\gamma$ used in this section, $a_{0}\simeq 1.72$, $x_{0}\simeq 1.07$, and $v_{0}\simeq 1.25$. When two extrema in Fig.~\ref{fig2} are about
to coincide Eq.~(\ref{103}) reads
\begin{equation} 
\label{103e}
x-x_{0}+0.24iy=0.96(v-v_{0})^{3}-0.90(a-a_{0})(v-v_{0}).
\end{equation}
The singular part of a solution of the cubic equation (\ref{103e}) can be evaluated as
\begin{equation} 
\label{103f}
(v-v_{0})\sim\sqrt{x-x_{0}+0.24i(y-\Delta)},
\end{equation}
where
\begin{equation} 
\label{103g}
\Delta\simeq 1.43(a_{0}-a)^{3/2}.
\end{equation}
According to Eq.~(\ref{9}), the part (\ref{103f}) contributes to $\partial\sigma/\partial x$. Therefore at $a_{0}<a$
\begin{eqnarray} 
\label{103h}
&&i\sigma(x,y)-i\sigma(x_{0}+0.24|\Delta|,0)\\
\nonumber
&&\sim\left[x-x_{0}-0.24|\Delta| +0.24iy\right]^{3/2}
\end{eqnarray}
and at $a<a_{0}$
\begin{equation} 
\label{103i}
i\sigma(x,y)-i\sigma(x_{0},\Delta)\sim\left[x-x_{0}+0.24i(y-\Delta)\right]^{3/2}.
\end{equation}
The singular part of the action (\ref{103h}) corresponds to Eq.~(\ref{103aa}) since $x_{c_{2}}=x_{0}+0.24|\Delta|$. The positions of 
singularities in the $\{x,y\}$ plane are
\begin{eqnarray}
\label{103j}
&&\{x_{0}\pm 0.24|\Delta|,\,0\},\hspace{0.3cm}a_{0}<a\\
\label{103l}
&&\{x_{0},\,\pm 0.24\Delta\},\hspace{0.3cm}a<a_{0}.
\end{eqnarray}
The singularities (\ref{103j}) relate to Fig.~\ref{fig3}(a) and ones (\ref{103l}) relate to Fig.~\ref{fig4}(b). At $a<a_{0}$ branches 1 and
3 are shown in Fig.~\ref{fig4}. At $y=\Delta$ the branch 1 touches the branch 3 at the point $c_{1}$ resulting in the singularity 
(\ref{103i}). Analogous singular point $c_{2}$  corresponds to $y=-\Delta$. At $-\Delta<y<\Delta$ reconnection occurs when the branches 1-3 
and 3-1 are formed, Fig.~\ref{fig4}(a). At $\Delta<|y|$ the branch 3 detouches and its top moves as a classical particle to infinity 
analogously to 
Fig.~\ref{fig3}(b).
\section{SEMICLASSICAL APPROXIMATION IN TWO DIMENSIONS}
\label{2D}
We use above the semiclassical formalism to solve the Schr\"{o}dinger equation (\ref{2}). This is possible when the semiclassical parameter 
$B$ is large. In this case one can reduce the exact equation (\ref{52}) to the Hamilton-Jacobi type (\ref{5}) and to obtain the wave function
(\ref{4}) in the exponential approximation when the action is large. To determine a prefactor one should go to next orders in the small 
parameter $1/B$ in Eq.~(\ref{52}). As a result, the preexponential factor is an asymptotic expansion with respect to $1/B$. 

At the first sight, the obtained solution of the Hamilton-Jacobi equation provides the main (exponential) part of the exact wave function 
with no problems. However, the prefactor may jump down to zero as a function of coordinates. This is called Stokes phenomenon 
\cite{STOKES,HEAD,MASL} considered in this section. 
\subsection{Semiclassical approximation in one dimension}                                         
To demonstrate the nature of Stokes phenomenon it is better to start with a one-dimensional static Schr\"{o}dinger equation. We consider a
known problem of overbarrier reflection when a particle energy $E$ is larger than a barrier hight $V$. To be specific we choose the 
potential barrier in the form $V(x)=V/\cosh^{2}(x/a)$. In classical mechanics overbarrier reflection is forbidden. In quantum mechanics, 
generally speaking, a reflected wave is not zero. The incident wave in WKB approximation is
\begin{equation} 
\label{201}
\psi_{I}\sim\exp\left(ik\int^{x}_{0}dx_{1}\sqrt{1-\frac{V(x_{1})}{E}}\right),
\end{equation}
where $k=\sqrt{2mE}/\hbar$ is a wave vector at the infinity \cite{LANDAU}. 
\begin{figure}
\includegraphics[width=4.5cm]{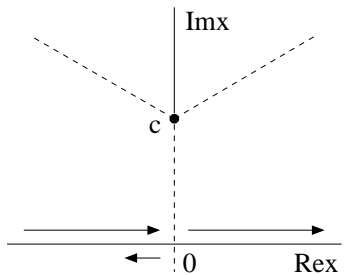}
\caption{\label{fig5}Stokes lines in the complex plane of $x$ in the case of a one-dimensional overbarrier reflection. Long arrows show an
incident wave. The short arrow indicates the reflected wave existing at negative $x$ only and resulted from Stokes phenomenon.} 
\end{figure}

The solution (\ref{201}) holds at every $x$ since the momentum $\sqrt{2m}\sqrt{E-V(x)}$ is nowhere zero. This put a reasonable question,
how and where a reflected wave forms. To answer the question one should consider the whole complex plane of $x$. A terminal point, where 
$E=V(x)$, is imaginary, $x_{c}=ia\arctan\sqrt{E/V-1}$. The exact solution of the Schr\"{o}dinger equation with the potential $V(x)$ can be 
written in the form
\begin{equation} 
\label{202}
\psi=\exp\left[\varphi(x)\right].
\end{equation}
In the semiclassical approximation close to the terminal point 
\begin{equation} 
\label{203}
\varphi(x)-\varphi(x_{c})\simeq\frac{ka}{3}\left[\frac{2i(x-x_{c})}{a}\right]^{3/2}\left(\frac{E-V}{V}\right)^{1/4}.
\end{equation}
The terminal point, marked in Fig.~\ref{fig5} as $c$, is an origen of Stokes lines which are determined by the condition 
${\rm Im[\varphi(x)]=0}$ and are shown by dashed lines. The point $c$ is a branching point. Along a Stokes line an increasing (dominant) 
solution of the Schr\"{o}dinger equation increases most fast and a decreasing (subdominant) solution falls most fast compared to neighbor 
directions from the point $c$. According to this definition, the term ``subdominant'' is applicable to a branch even when it is a single one.

A dominant branch is continuous when we cross a Stokes line. If on some Stokes line a dominant branch is absent, a subdominant branch is 
also continuous after crossing the Stokes line. 

The case of coexistence on a Stokes line of two types of branches, dominant and subdominant, is not trivial. Strictly speaking, it is 
impossible to keep an exponentially small (subdominant) branch at those $x$ where an exponentially large (dominant) branch exists. As shown 
by Stokes \cite{STOKES}, in presence of a dominant branch a coefficient at a subdominant branch can jump after crossing a Stokes line. This 
is called Stokes phenomenon \cite{HEAD,MASL}. 

In our case of overbarrier reflection the incident wave (\ref{201}) is indicated in Fig.~\ref{fig5} by long arrows. The exact solution 
(\ref{202}) at real $x$ is a sum of the incident (dominant) wave and a reflected (subdominant) one
\begin{equation} 
\label{204}
\psi(x)=\psi_{I}(x)+\psi_{R}(x),
\end{equation}
where in the semiclassical approximation the reflected wave is
\begin{eqnarray} 
\label{205}
\psi_{R}(x)\sim\theta(-x)\exp\left[-\pi ak\left(1-\sqrt{V/E}\right)\right]\\
\nonumber
\exp\left(-ik\int^{x}_{0}dx_{1}\sqrt{1-\frac{V(x_{1})}{E}}\right).
\end{eqnarray}
The subdominant branch $\psi_{R}$, indicated by the short arrow in Fig.~\ref{fig5}, jumps to zero at positive $x$. This Stokes jump happens 
at the vertical Stokes line in Fig.~\ref{fig5}. 

A rapid jump of a prefactor, as in Eq.~(\ref{205}), occurs in a semiclassical solution only. An exact solution is continuous. To show that 
one should go beyond the semiclassical approximation considering a ``microstructure'' of a Stokes line. 
\begin{figure}
\includegraphics[width=5.5cm]{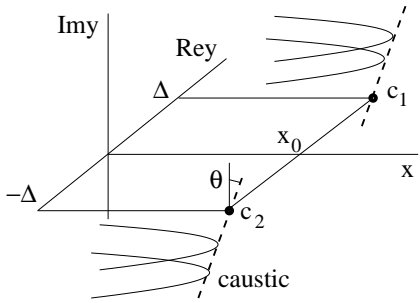}
\caption{\label{fig6}Classical trajectories are reflected from caustics marked by the dashed lines which pierce the physical plane 
$\{x,{\rm Re}y\}$ at the points $\{x_{0},\pm\Delta\}$. The trajectories and the caustics lie in the planes ${\rm Re}y=\pm\Delta$. The both
caustics are tilted by the angle $\theta$ with respect to the vertical direction. The point $c_{1}$ is the same as in Fig.~\ref{fig4}(b).} 
\end{figure}
\subsection{Two dimensions.}
When dimensionality is more than one, semiclassical approximation becomes much more complicated \cite{MASL}. In this paper we do not concern 
general aspects of this problem. We only analyze the semiclassical solution of Sec.~\ref{sol}. 

First, there is an analogy in properties of wave functions in one and two dimensions. In one dimension the wave function has the branching
point (\ref{203}) in the plane $\{{\rm Re}x,\,{\rm Im}x\}$. In two dimensions a similar branching point (\ref{103i}) is in the plane 
$\{x,\,{\rm Re}y\}$. 

The singularity in the $\{x,\,{\rm Re}y\}$ plane can be studied by extention of dimensionality to three dimensions, 
$\{x,\,{\rm Re}y,\,{\rm Im}y\}$, as shown in Fig.~\ref{fig6}. To do that one can consider the action (\ref{103i}) close to the point $c_{1}$
in the plane ${\rm Re}y=\Delta$ in Fig.~\ref{fig6}, that is $\sigma(x,\Delta+i\eta)$. In the plane ${\rm Re}y=\Delta$ classical trajectories
are reflected from the certain lines as shown in Fig.~\ref{fig6}. Those lines, called caustics \cite{LANDAU1,MASL}, came to quantum 
mechanics from optics. Underbarrier caustics were studied in Refs.~\cite{SCHMID1,SCHMID2,DYK}. The caustics pierce the physical plane 
$\{x,\,{\rm Re}y\}$ at the points $\{x_{0},\,\pm\Delta\}$. The classical trajectories, reflected from the caustics, are
\begin{eqnarray} 
\label{206}
&&x(\eta,b)=\frac{x_{0}+(1-\gamma)\tan^{2}\theta+b\tan\theta}{1+\tan^{2}\theta}\\
\nonumber
&&-\frac{(1+\tan^{2}\theta)(\eta-b)^{2}}{4(x_{0}-1+\gamma+b\tan\theta)}.
\end{eqnarray}
The parameter $b$ marks different trajectories in Fig.~\ref{fig6}. It is not difficult to check that the trajectory (\ref{206}) relates to 
the energy $\gamma-1$. The relation $\partial x/\partial\eta=\tan\theta$ determines at each $b$ the certain point on the trajectory where it
is tangent to the caustic. Each caustic is defined by the relation $x-x_{0}=\eta\tan\theta$. The caustics are straight lines in small 
vicinities of the points $c_{1}$ and $c_{2}$ in Fig.~\ref{fig6}. 

As in a one-dimensional case, the singular part of the action depends on a distance, to the power 3/2, in the direction perpendicular to 
the caustic, that is on $[(x-x_{0})\cos\eta-\eta\sin\eta]^{3/2}$. Comparing this with the expression (\ref{103i}), where $y=\Delta+i\eta$, 
one can conclude that $\tan\theta=0.24$ and therefore in Fig.~\ref{fig6} $\theta\simeq 13.5^{o}$ for the parameters chosen in 
Sec.~\ref{sol}. 
\begin{figure}
\includegraphics[width=4cm]{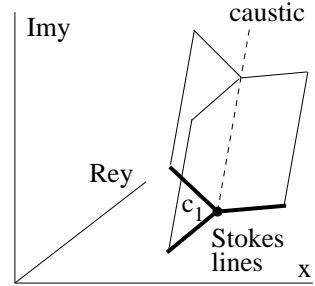}
\caption{\label{fig7} The caustic (the dashed line) is an origen of Stokes planes which intersect the physical plane $\{x,\,{\rm Re}y\}$ 
along Stokes lines.} 
\end{figure}

In a vicinity of each point of the caustic the semiclassical action is singular having a branching point of the type (\ref{103i}). The 
caustic is an origen of Stokes planes shown in Fig.~\ref{fig7}. The Stokes planes in the three dimensional space are analogous to Stokes 
lines in a plane. The Stokes planes intersect the physical plane $\{x,\,{\rm Re}y\}$ along the Stokes lines plotted in Fig.~\ref{fig7}. 

Now let us move along the branch 1-3 in Fig.~\ref{fig4}(a) from $x=0$. This branch remains dominant when $x$ is less than the intersection 
point with the branch 3-1. Then it becomes subdominant within the finite interval of $x$. According to Stokes theory \cite{HEAD}, at some
point within that interval a jump of the branch 1-3 down to zero is possible. Therefore there is no guarantee that at a larger $x$, when 
the branch 1-3 becomes dominant again, it would be the same as a formal continuation of 1-3 according to semiclassical formulas.
 
To analyze that situation we consider the case when the parameter $a$ is close to $a_{0}$ defined in Sec.~\ref{sol}. This means that the
points $c_{1}$ and $c_{2}$ in Fig.~\ref{fig6} are close to each other. At parameters chosen in Sec.~\ref{sol}, Eq.~(\ref{103e}) at $y=0$ 
yields 
\begin{equation} 
\label{207}
\frac{\partial x}{\partial v}=2.88\left[(v-v_{0}-\delta)^{2}+2(v-v_{0}-\delta)\delta\right],
\end{equation}
where $\delta=0.56\sqrt{a-a_{0}}$ is a small parameter. The value $v=v_{0}+\delta$, at $a_{0}<a$, corresponds to the singular point $c_{2}$ 
in Figs.~\ref{fig2} and \ref{fig3}(a). According to Eq.~(\ref{9}), at $x$ close to $x_{0}$ and $v$ close to $v_{0}$ one can write at $y=0$
\begin{equation} 
\label{208}
\frac{\partial i\sigma}{\partial x}=0.13+0.20(v-v_{0}-\delta). 
\end{equation}
Equation (\ref{207}) allows to integrate in Eq.~(\ref{208}). The result at $a<a_{0}$ is
\begin{eqnarray} 
\label{209}
&&i\sigma(x,0)-i\sigma(x_{0},0)-0.13(x-x_{0})\\
\nonumber
&&=0.14\left[(v-v_{0}-i|\delta|)^{4}+\frac{8i}{3}(v-v_{0}-i|\delta|)^{3}|\delta|\right].       
\end{eqnarray}
The right hand side of Eq.~(\ref{209}) is a singular part of the action in terms of the variable $v$. According to Eq.~(\ref{207}), a small 
$(v-v_{0}-\delta)$ is proportional to a square root of the distance and therefore the cubic part of the expression (\ref{209}) has the same 
type of singularity as Eqs.~(\ref{103h}) and (\ref{103i}). Another singular part of the action can be obtained from Eq.~(\ref{209}) by the
formal change $|\delta|\rightarrow -|\delta|$. 
\begin{figure}
\includegraphics[width=4cm]{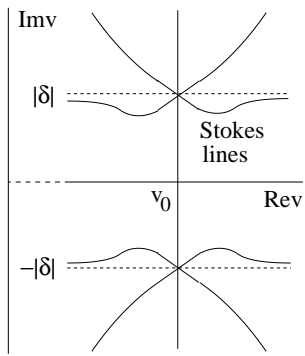}
\caption{\label{fig8}Stokes lines of the action $\sigma(x,0)$ which is considered as a function of the complex variable $v$ at $a<a_{0}$.}
\end{figure}

Below we consider the case of $a<a_{0}$. The singular part of the action, the right hand side of Eq.~(\ref{209}), is a function of the 
complex variable $v$. This part becomes zero at the point $v=v_{0}+i|\delta|$ which is an origen of Stokes lines in the complex plane of the 
variable $v$. Those lines are determined by the condition 
\begin{equation} 
\label{210}
{\rm Im}\left[(v-v_{0}-i|\delta|)^{4}+\frac{8i}{3}(v-v_{0}-i|\delta|)^{3}|\delta|\right]=0
\end{equation}
and are shown in Fig.~\ref{fig8}. 

There are three branches in the problem indicated in Fig.~\ref{fig3} or Fig.~\ref{fig4}. The branch 1-3 in Fig.~\ref{fig4}(a) for all $x$ 
relates to real $v$ in Eq.~(\ref{103}) but the branches 1-3 and 2-2 are determined by $v$ with finite imaginary parts. This means that on 
the line ${\rm Im}v=0$ in Fig.~\ref{fig8} there is only one branch which is 1-3 of Fig.~\ref{fig4}(a). We see that different branches are
localized at different parts of the complex plane of $v$. Hence they cannot be classified as dominant and subdominant. In contrast, in 
terms of conventional coordinates all branches exist at the same $x$ where dominant-subdominant competition is unclear. 

Motion along the branch 1-3 in Fig.~\ref{fig4}(a) from $x=0$ is equivalent to sliding along the line ${\rm Im}v=0$ in Fig.~\ref{fig8} from
$v=0$.  According to Eq.~(\ref{209}), in a small vicinity of $v_{0}$ the single (at real $v$) branch becomes subdominant. Obviously, it is 
continuous after crossing the Stokes line at the point $v_{0}$ in Fig.~\ref{fig8}.  

One can conclude from here that despite the branch 1-3 in Fig.~\ref{fig4}(a) becomes subdominant within the certain domain of $x$ there are 
no Stokes jumps there and the branch 1-3 goes continuously through that domain according to the semiclassical formulas obtained.

In vicinities of the points $a$ and $b$ in Figs.~\ref{fig3} and \ref{fig4} the imaginary part of the action is quadratic with respect to 
$(x-x_{a,b})$. In a one-dimensional case an action also can be quadratic when $V(x)-E$ is quadratic close to some point $x$. This point is
also an origen of Stokes lines in the one-dimensional case \cite{HEAD}. 

In two dimensions the extremum of ${\rm Im}\sigma$ at the point $b$ is continued to the whole classical trajectory (\ref{103d}). The 
component of the vector $\nabla\sigma$, parallel to the trajectory, is continuous if to cross it and is determined by Eqs.~(\ref{9}) and 
(\ref{10}). Therefore the perpendicular component of $\nabla\sigma$, as a part of the whole vector $\nabla\sigma$, is also continuous in 
crossing the trajectory. The same is valid for the point  $a$. It follows that in vicinities of the points $a$ and $b$ the branches are 
determined by the semiclassical solution and look as ones shown in Figs.~\ref{fig3} and \ref{fig4}. 

So we proved that the semiclassical solution of Secs.~\ref{ham-jac} and \ref{sol} determines the exact wave function with exponential 
accuracy.  
\section{CLASSICAL TRAJECTORIES IN IMAGINARY TIME}
\label{traj}
A particle penetrates through the barrier from the $\delta$ well to the outer region where it moves along the classical trajectory 
(\ref{103d}). Accordingly, a barrier penetration parameter can be defined as 
\begin{equation} 
\label{211}
w=\bigg|\frac{\psi(x_{b},0)}{\psi(0,0)}\bigg|^{2}.
\end{equation}
In Sec.~\ref{dyn} we discuss a meaning of the parameter $w$. In the semiclassical approximation $|\psi(0,y)|$ does not depend on $y$ and 
$|\psi(x,y)|$ on the trajectory (\ref{103d}) is also a constant. $\{x_{b},0)\}$ is just a point of the trajectory (\ref{103d}). By means of 
Eq.~(\ref{4}) the parameter (\ref{211}), with the exponential accuracy, is
\begin{equation} 
\label{17}
w\sim\exp\left\{-2B\,{\rm Im}\left[\sigma(x_{b},0)-\sigma(0,0)\right]\right\}.
\end{equation}
To find $w$ one can substitute the solution of Sec.~\ref{sol} into Eq.~(\ref{17}). 

But there is another way. For a calculation of $w$ it is not necessary to know the action in the entire $\{x,y\}$ plane. If we know the 
action on a certain path $x(y)$, which connects two points $\{x_{b},0\}$ and $\{0,0\}$, it is sufficient to determine $w$. A role of such 
path can be played by a classical trajectory in imaginary time since a motion under a barrier in real time is classically forbidden. In the 
method of classical complex trajectories in imaginary time $t=i\tau$ the coordinate $x(\tau)$ remains real but the other coordinate becomes 
imaginary $y(\tau)=i\eta(\tau)$. This type of complex coordinates was used in magnetotunneling \cite{BLATT}. See also 
Refs.~\cite{DYK,GOROKH}.

The Hamilton-Jacobi equation (\ref{5}), according to rules of classical mechanics \cite{LANDAU2}, generates the equations of motion 
\begin{equation} 
\label{21}
\frac{1}{2}\frac{\partial^{2}x}{\partial\tau^{2}}=-1,\hspace{0.5cm}
\frac{1}{2}\frac{\partial^{2}\eta}{\partial\tau^{2}}=0,
\end{equation}
where time is measured in the units of $\hbar B/u_{0}$. The boundary conditions to Eqs.~(\ref{21})
\begin{eqnarray} 
\label{21a}
&&x(0)=x_{b},\hspace{0.5cm}\frac{\partial x}{\partial\tau}\bigg|_{0}=0,\\
\nonumber
&&\eta(0)=0,\hspace{0.5cm}\frac{\partial\eta}{\partial\tau}\bigg|_{0}=2\sqrt{x_{b}-1+\gamma}
\end{eqnarray}
are compatible with the energy conservation 
\begin{equation} 
\label{109}
-\frac{1}{4}\left(\frac{\partial x}{\partial\tau}\right)^{2}+\frac{1}{4}\left(\frac{\partial\eta}{\partial\tau}\right)^{2}-x=\gamma-1.
\end{equation}
The trajectory
\begin{equation} 
\label{22}
x(\tau)=x_{b}-\tau^{2},\hspace{0.3cm}\eta(\tau)=2\tau\sqrt{x_{b}-1+\gamma}
\end{equation}
is a solution of Eq.~(\ref{21}). It starts at the point $\tau=0$ (the exit point from under the barrier) and terminates at 
$\tau_{0}=\sqrt{x_{b}}$ when $x(\tau_{0})=0$ (the position of the $\delta$ well). The terminal value 
$\eta(\tau_{0})=2\sqrt{x_{b}(x_{b}-1+\gamma)}$ coincides with $v_{b}$ as follows from Eqs.~(\ref{103b}) and (\ref{103c}). It also follows 
from Eq.~(\ref{109}) that the relation 
\begin{equation} 
\label{22a}
\frac{\partial x}{\partial\tau}\bigg|_{\tau_{0}}=-2\sqrt{1+\alpha^{2}(iv_{b})},
\end{equation}
holds which corresponds to the condition (\ref{101}) at the $\delta$ well. 

\begin{figure}
\includegraphics[width=5cm]{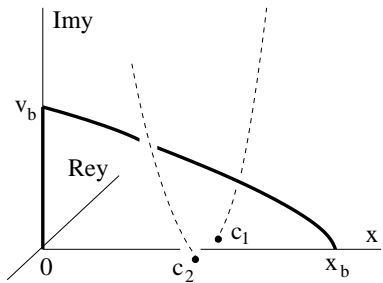}
\caption{\label{fig9}The bypass in the plane ${\rm Re}y=0$ is the solid curve along which the action is tracked. The part, connecting the 
points $\{0,iv_{b}\}$ and $\{x_{b},0\}$, relates to the classical trajectory in imaginary time. The caustics are shown by the dashed curves.}
\end{figure}

The trajectory (\ref{22}) is shown in Fig.~\ref{fig9} as the curve connecting the points $\{x_{b},0\}$ and $\{0,iv_{b}\}$. The barrier 
penetration parameter (\ref{17}) can be written in the form
\begin{equation}
\label{23a}
w\sim\exp\left(-A_{0}-A_{1}\right),
\end{equation}
where
\begin{equation} 
\label{19}
{A}_{0}=2B\,{\rm Im}\left[\sigma(x_{b},0)-\sigma(0,iv_{b})\right]
\end{equation}
and
\begin{equation} 
\label{23}
A_{1}=2B{\rm Im}\left[\sigma(0,iv_{b})-\sigma(0,0)\right].
\end{equation}

By means of the classical trajectory (\ref{22}) one can calculate the part $A_{0}$ as an integral of the Lagrangian with respect to time
\cite{LANDAU2} 
\begin{equation} 
\label{20}
A_{0}=2B\int^{\tau_{0}}_{0}\left[\frac{1}{4}\left(\frac{\partial x}{\partial\tau}\right)^{2}-
\frac{1}{4}\left(\frac{\partial\eta}{\partial\tau}\right)^{2}-x+1-\gamma\right]d\tau.
\end{equation}
With the trajectory (\ref{22}) Eq.~(\ref{20}) reads
\begin{equation} 
\label{20a}
A_{0}=\frac{4B}{3}\sqrt{1+\alpha^{2}(iv_{b})}\left[1-3\gamma-2\alpha^{2}(iv_{b})\right].
\end{equation}
The part $A_{1}$ cannot be determined by trajectories. One can use directly the Hamilton-Jacobi equation (\ref{5}) at $x=0$ with the
condition (\ref{101}). One should integrate the term $\partial\sigma/\partial y$. The path, connecting the points $\{0,iv_{b}\}$ and 
$\{0,0\}$, is the vertical solid line in Fig.~\ref{fig9}. After the integration we obtain 
\begin{equation} 
\label{24}
A_{1}=2B\int^{v_{b}}_{0}d\eta\sqrt{\gamma+\alpha^{2}(i\eta)}.
\end{equation}

On the other hand, the same result for $A_{0}+A_{1}$ follows from the direct solution (\ref{7}) of the Hamilton-Jacobi equation integrating 
the last term by parts and using the relation (\ref{103c}).

One should note that in the case of a homogeneous (in the $y$ direction) wire Eq.~(\ref{23a}) yields the conventional WKB result for
tunneling through a triangular barrier \cite{LANDAU} 
\begin{equation} 
\label{24a}
w\sim\exp\left(-\frac{4B}{3}\right),\hspace{0.3cm}\alpha(y)=0.
\end{equation}

The whole bypass, connecting the points $\{x_{b},0\}$ and $\{0,0\}$, is shown in Fig.~\ref{fig9}. It goes in the plane
${\rm Re}y=0$ between the two caustics which are continuations of ones from Fig.~\ref{fig6}. These caustics are origins of Stokes planes as
in Fig.~\ref{fig7}. One plane starts at the caustic $c_{1}$ and terminates at $c_{2}$. Therefore the path in Fig.~\ref{fig9} pierces that 
Stokes plane. According to Sec.~\ref{2D}, one can track the semiclassical solution along the path in Fig.~\ref{fig9} with no Stokes jumps. 

The conventional (no tangent momentum) underbarrier trajectory in Fig.~\ref{fig1}(a) corresponds to real coordinates. In contrast, the path
in Fig.~\ref{fig9} lies in the complex plane providing a real tangent velocity $\partial y/\partial t=\partial\eta/\partial\tau$. This path
can be treated as a saddle point with fluctuation around it. Those fluctuating paths in the real plane $\{x, {\rm Re}y\}$ look as a wide 
bundle of trajectories of the type shown in Fig.~\ref{fig1}(b).
\section{ABOUT A NUMERICAL CALCULATION}
\label{num}
In principle, the Schr\"{o}dinger equation (\ref{1}) can be solved numerically using a discrete two-dimensional net with small finite steps
$\Delta x$ and $\Delta y$. The error in use of such numerical scheme is of the order of $(\Delta x)^{2}\sim(\Delta y)^{2}$. A reduction of 
steps increases a precision of numerical calculations. 

A specificity of the problem is that one should keep at the same spatial domain two branches of the wave function, dominant and subdominant. 
In Fig.~\ref{fig3}(b) the branches 1 and 2 merge when, according to Eq.~(\ref{24a}), $|\psi|\sim\exp\left(-2B/3\right)$. The subdominant 
branch 2 at $x=0$ in Fig.~\ref{fig3}(b) can be estimated as $|\psi|\sim\exp\left(-4B/3\right)$. 

To resolve the exponentially small subdominant branch the precision of numerical calculations should be no lesser than 
$(\Delta x)^{2}\sim\exp\left(-4B/3\right)$. So the number $N\sim1/\Delta x\Delta y$ of points in the two-dimensional discrete net should be 
no smaller than $\exp\left(4B/3\right)$. 

The parameter $B$ has to be sufficiently large to get a developed semiclassical regime. On the other hand, $B$ should not be too large
since it results in a nonrealistically large number of discrete points in the net. A choice of $B$ for numerical calculations is a compromise
between the two tendencies. 

According to our experience in numerical studies of this type of problems \cite{IVLEV8}, a semiclassical regime occurs when $B$ is no 
smaller than approximately 25. In this case the number of points $N$ should be no smaller than $10^{14}$ which is essentially outside of 
memory facilities of computers. This makes impossible a straightforward numerical investigation of the problem. 
\section{HOW A PARTICLE PENETRATES THROUGH THE BARRIER}
\label{dyn}
We specify the same parameters $\gamma=0.2$ and $\alpha^{2}_{0}=0.03$ as in Sec.~\ref{sol}. One can use the direct solution of the
Hamilton-Jacobi equation, Sec.~\ref{sol}, or trajectory results, Sec.~\ref{traj}, to calculate the barrier penetration parameter (\ref{211})
which, at  a small $|a_{R}-a|$, has the form
\begin{equation} 
\label{212}
w\sim\exp\left[-2.0B(a_{R}-a)\right],
\end{equation}
where $a_{R}\simeq 2.27$. In dimension units Eq.~(\ref{212}) reads
\begin{equation} 
\label{213}
w\sim\exp\left[-4.54B\left(1-\frac{{\cal E}}{{\cal E}_{R}}\right)\right],
\end{equation}
where ${\cal E}_{R}\simeq 2.27u_{0}/a$. The result (\ref{213}) holds when $|1-{\cal E}/{\cal E}_{R}|$ is small. As follows, when the 
electric field is close to ${\cal E}_{R}$ the parameter $w$ becomes not exponentially small. In this case in Fig.~\ref{fig3}(b) 
$x_{b}\simeq 2.0$. One should note that tunneling rate at ${\cal E}={\cal E}_{R}$ from a homogeneous wire, $\alpha(y)=0$, is exponentially 
small according to WKB. 

The barrier penetration parameter $w$ is rather formal. To physically formulate a problem of barrier penetration one should localize a 
particle in a vicinity of the $\delta$ well and than study a subsequent dynamics. Suppose that the wave function, which is localized at the 
$\delta$ well at $t=0$, is $\varphi(\vec r)$, where $\vec r=\{x,y\}$. The function $\varphi(\vec r)$ coincides with $\psi(\vec r)$ in 
Fig.~\ref{fig4} at $x<x_{a}$ and $\varphi(\vec r)=0$ at a larger $x$. 

The system, being released at $t=0$, starts up with the abrupt function $\varphi(\vec r)$ to restore the true wave function. In the 
conventional case of a homogeneous wire, $\alpha(y)=0$, during the short time $\hbar/u_{0}$ the branches 1 and 2, as in Fig.~\ref{fig3}(b), 
are formed. Then the process becomes exponentially slow when a weak leakage through the barrier is provided by an outgoing flux related to
the right hand side of the branches 1 and 2. This process can be described in terms of an exponentially small imaginary part of the total 
energy.

When the wire is not homogeneous, $\alpha(y)\neq0$, a scenario becomes different. The function $\psi(\vec r)$, plotted in Fig.~\ref{fig4} is
almost an eigenfunction with a real energy if to ignore an exponentially weak leakage across the barrier provided by the right hand side of 
branches 1 and 2. Any local artificial distortion of the eigenfunction is restored during a time interval inversely proportional to the
barrier height. This is similar to the case of a homogeneous wire. The use of the abrupt function $\varphi(\vec r)$ instead
of exact $\psi(\vec r)$ is an example of such distortion. Therefore a conversion of $\varphi(\vec r)$ into $\psi(\vec r)$ is characterized 
by the time $\hbar/u_{0}$. Within the short (nonsemiclassical) time $\hbar/u_{0}$, due to the uncertainty principle, states with all 
energies are involved including overbarrier propagating waves. These waves provide a probability transfer from the $\delta$ well to the 
outer region.
\begin{figure}
\includegraphics[width=4.5cm]{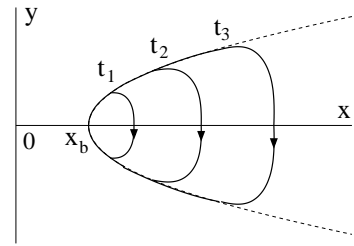}
\caption{\label{fig10}The scheme of an expansion in time of the state which tends to restore the true wave function localized along the 
classical trajectory shown by the dashed curve.} 
\end{figure}

The subsequent dynamics is schematically shown in Fig.~\ref{fig10} where the expanding in time probability density tends to restore 
$\psi(\vec r)$ localized on the classical trajectory (\ref{103d}). This expansion occurs infinitely if the space $x_{b}<x$ is not restricted.

At a relatively small electric field, ${\cal E}<{\cal E}_{R}$, the particle density on the trajectory is exponentially small. The decay 
rate (filling out of the trajectory in Fig.~\ref{fig10}) of the state at the $\delta$ well is exponentially weak and is determined by $w$, 
Eq.~(\ref{213}). We do not consider a preexponential factor having the dimensionality of inverse time. 

At a larger electric field, ${\cal E}_{R}<{\cal E}$, the peak of density outside the barrier, at $x=x_{b}$, is exponentially large compared 
to one at the $\delta$ well as follows from Eq.~(\ref{213}). This means that the region at the $\delta$ well is emptied, down to an 
exponentially small density, during the short time scale $\hbar/u_{0}$. 

So one can describe the phenomenon as follows. Suppose a state in the non-homogeneous wire to be not a ground state. Under increase of the 
electric field, above the certain threshold ${\cal E}_{R}$, the prebarrier region gets emptied fast (instant tunneling). The threshold
${\cal E}_{R}$ depends on energy of a decaying state in the quantum wire. For a higher energy ${\cal E}_{R}$ is less.
\section{TUNNELING THROUGH A BARRIER WITH AN IMPURITY}
\label{imp}
\begin{figure}
\includegraphics[width=5.0cm]{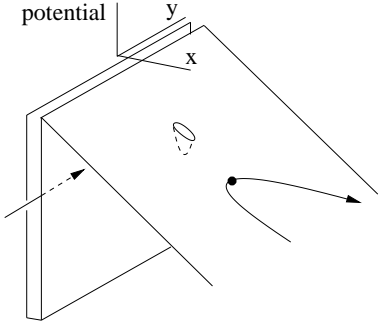}
\caption{\label{fig11}The potential barrier with the impurity, related to Eq.~(\ref{217}), is plotted for positive $x$. The state with a 
finite prebarrier momentum is shown by the arrow line. The state after the barrier is indicated as the classical trajectory reflected from 
the barrier at the point $x\simeq 1$, $y=0$ marked by the dot.} 
\end{figure}

In this section we consider tunneling from a homogeneous quantum wire described in Sec.~\ref{form} where one has to put $\alpha(y)=0$. But
now the barrier is non-homogeneous in the direction parallel to the wire. Namely, there is an impurity at the barrier region sketched in 
Fig.~\ref{fig11}. In the dimensionless units of Sec.~\ref{finite} the Schr\"{o}dinger equation has the form
\begin{equation} 
\label{217}
-\frac{1}{B^{2}}\left(\frac{\partial^{2}\psi}{\partial x^{2}}+\frac{\partial^{2}\psi}{\partial y^{2}}\right)+\left[V_{0}(x)+u(x,y)\right]\psi
=(\gamma-1)\psi,
\end{equation}
where $u(x,y)$ is a potential of the impurity and
\begin{equation} 
\label{218}
V_{0}=-\frac{2}{B}\,\delta(x)-|x|.
\end{equation}
For simplicity we consider a symmetric impurity potential $u(-x,y)=u(x,y)$. This is 
equivalent to two identical impurities symmetrically localized around the wire.
\subsection{Hamilton-Jacobi approach}
\label{ham1}
We apply a semiclassical method based on the Hamilton-Jacobi equation to tunneling through a barrier with an impurity. The wave function has 
the form (\ref{4}) resulting in the Hamilton-Jacobi equation for positive $x$
\begin{equation} 
\label{215}
\left(\frac{\partial\sigma}{\partial x}\right)^{2}+\left(\frac{\partial\sigma}{\partial y}\right)^{2}-x+u(x,y)=\gamma-1.
\end{equation}

The equation (\ref{215}) should be complemented by a boundary condition at $x=0$. The function $\sigma(x,y)$ is continuous on the line $x=0$
and $(\partial\sigma/\partial x)^{2}$ has the same values at $x=\pm 0$. Since the total potential is symmetric with respect to $x$ the 
boundary condition at $x=+0$, accounting for the $\delta$ function in Eq.~(\ref{217}), is
\begin{equation} 
\label{216}
\frac{\partial\sigma(x,y)}{\partial x}\bigg|_{x=0}=i.
\end{equation}
We consider in this section a weak impurity, $u(x,y)\ll 1$, which enables to treat it as a perturbation in the Hamilton-Jacobi 
equation. We emphasize that the possibility to consider $u(x,y)$ as a conventional perturbation in the Schr\"{o}dinger equation (\ref{217}) 
is realized under more rigorous condition $u(x,y)\ll 1/B$. 

If to neglect $u(x,y)$ the solution of the Hamilton-Jacobi equation is
\begin{equation} 
\label{54}
\sigma_{0}(x,y)=ky+i\int^{x}_{0}dx_{1}\sqrt{1-x_{1}},\hspace{0.5cm}\gamma=k^{2},
\end{equation}
where the real parameter $k$ has a meaning of a wave vector parallel to the wire. At a finite $k$ the solution (\ref{54}) is not a ground 
state and corresponds to the finite underbarrier phase $ky$. This is an essential feature of the underbarrier state. The state with a finite 
momentum in the $\delta$ well is indicated in Fig.~\ref{fig11} as the arrow line. Since we consider an underbarrier state with a negative 
energy, it should be $k^{2}<1$. Strictly speaking, $\gamma$ contains an exponentially small imaginary part related to decay of 
the metastable state at the $\delta$ well. We omit that small correction in calculations of the wave function.

In the next order with respect to $u(x,y)$ the solution can be written in the form
\begin{equation} 
\label{55}
\sigma=\sigma_{0}(x,y)+\sigma_{1}(x,y),\hspace{0.5cm}\frac{\partial\sigma_{1}(x,y)}{\partial x}\bigg|_{x=0}=0.
\end{equation}
The correction $\sigma_{1}$ satisfies the equation
\begin{equation} 
\label{56}
2i\frac{\partial\sigma_{1}}{\partial x}\sqrt{1-x}+2k\frac{\partial\sigma_{1}}{\partial y}=-u(x,y).
\end{equation}
The solution of Eq.~(\ref{56}), obeying the boundary condition (\ref{55}), is
\begin{align} 
\label{57}
\sigma_{1}(x,y)=\int^{\infty}_{0}\frac{dy_{1}}{2k}u\left(0,y_{1}+y-2ik\sqrt{1-x}+2ik\right)\\
\nonumber
+\int^{x}_{0}\frac{idx_{1}}{2\sqrt{1-x_{1}}}u\left(x_{1},y-2ik\sqrt{1-x}+2ik\sqrt{1-x_{1}}\right).
\end{align}
The correction $\sigma_{1}$ to the action describes scattering of underbarrier waves by the impurity.
\subsection{Solution}
\label{sol1}
Now one should specify a particular form of the impurity potential $u(x,y)$. We use the exponential form
\begin{align} 
\label{58}
u(x,y)=-u\exp\left[-\frac{(x-l)^{2}+y^{2}}{a^{2}}\right]\\
\nonumber
-u\exp\left[-\frac{(x+l)^{2}+y^{2}}{a^{2}}\right]
\end{align}
with a small parameter $a\ll 1$. The impurity potential is well localized. The dimensionless parameters $v_{0}$, $l$, and $a$ can be 
easily expressed through corresponding physical ones.

A simple analysis of Eq.~(\ref{57}) shows that outside the barrier a maximum of $|\psi(x,y)|$ is reached on the classical trajectory 
$y=2k\sqrt{x-1}$. In the semiclassical approximation used $|\psi(x,y)|$ is a constant along the trajectory. It smears out if to account for 
quantum effects described by the last terms in Eq.~(\ref{52}). That solution, localized at the classical trajectory, is symbolically shown 
in Fig.~\ref{fig11} by the arrow curve. 

The integration in Eq.~(\ref{57}) is not difficult. Under the condition 
\begin{equation} 
\label{58a}
l<2k^{2}<2
\end{equation}
the modulus of the wave function, at $1<x$ and close to the classical trajectory $y=2k\sqrt{x-1}$, has the form
\begin{align} 
\label{59}
|\psi(x,y)|=\frac{e^{-2B/3}}{(x-1)^{1/4}}\exp\Bigg\{\frac{Ba^{2}lu}{8k^{2}(2k^{2}-l)}\exp\left(\frac{4k^{2}-l^{2}}{a^{2}}\right)\\
\nonumber
\exp\left[-\frac{\left(y-2k\sqrt{x-1}\right)^{2}}{a^{2}}\right]\cos\left[\frac{4k}{a^{2}}\left(y-2k\sqrt{x-1}\right)\right]\Bigg\}.
\end{align}
At $u=0$ (no impurity) Eq.~(\ref{59}) turns into a conventional WKB expression. Eq.~(\ref{59}) describes the state outside the barrier 
indicated by the arrow curve in Fig.~\ref{fig4}. This state is generic with one at the right hand side part of Fig.~\ref{fig1}(b). The state
(\ref{59}), driven by underbarrier mechanisms, essentially differs from a state outside the barrier without the impurity (incident and 
reflected waves) which is hardly influenced by the $\delta$ well.

Far from the impurity, $|y|\rightarrow\infty$, and away from the classical trajectory the wave function coincides with the conventional WKB 
form which is the first factor in Eq.~(\ref{59}). In this case there is an outgoing wave only outside the barrier. It provides an 
exponentially small imaginary part of the energy which is neglected in Eq.~(\ref{59}). If to ignore the exponentially weak leakage across 
the barrier the wave function (\ref{59}) can be treated as an eigenfunction. 

Above we did not account for a small correction $\delta\gamma$ to the eigenvalue $\gamma-1$ (\ref{54}) due to the impurity 
potential. This real correction would result in the additional part $y\delta\gamma/2k$ in $\sigma_{1}(x,y)$ which does not influence 
the modulus of the wave function (\ref{59}).

The applicability conditions of the result (\ref{59}) are
\begin{equation} 
\label{60}
u\exp\left(\frac{4k^{2}-l^{2}}{a^{2}}\right)\ll 1,\hspace{0.3cm}\exp\left(\frac{4k^{2}-l^{2}}{a^{2}}\right)\ll B.
\end{equation}
We remind that $B$ is a large parameter. The first inequality (\ref{60}) follows from the perturbation condition with respect 
to the impurity potential. The second condition (\ref{60}) is semiclassical one when the last part in Eq.~(\ref{52}) is less 
than $u(x,y)$. 
\subsection{Results}
One can draw two conclusions on the basis of equation (\ref{59}): (i) the effective amplitude of the impurity potential 
$u\exp[(4k^{2}-l^{2})/a^{2}]$ is exponentially enhanced compared to $u$ since $k\sim l\sim 1$ and $a$ is small and (ii) a scenario of 
barrier penetration is of the type as in Fig.~\ref{fig1}(b).

The underbarrier exponential  enhancement is generic with one occurring in tunneling through nonstationary one-dimensional barriers 
\cite{IVLEV4} where there is an interference of various paths generated in different moments of time. 

In Fig.~\ref{fig11} the impurity position $l$ is before the exit point. According to the conditions (\ref{58a}), the impurity can be 
placed even after the exit point, $1<l$.
\begin{figure}
\includegraphics[width=4.5cm]{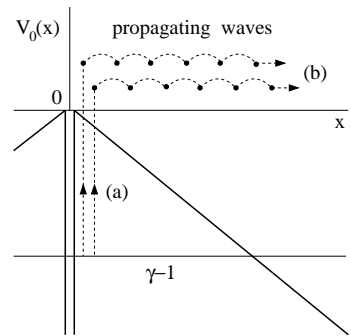}
\caption{\label{fig12}The processes of scattering by the impurity placed close to the position $x=0$. The eigenstate in the total potential
$V_{0}(x)+u(x,y)$ (the barrier plus the impurity) corresponds to the underbarrier eigenvalue of the energy $\gamma-1$. The propagating waves 
(eigenstates of $V_{0}(x)$) are a part of the total eigenstate. The dots mark impurity scattering.} 
\end{figure}
\section{THE NATURE OF THE PHENOMENON}
\label{nature}
In this section general arguments are formulated which allow to analyze the phenomenon by ``hand waving''. To be specific we consider 
tunneling from a homogeneous quantum wire through the barrier with an impurity (\ref{217}).

In absence of the impurity ($u(x,y)=0$) motions in $x$ and $y$ directions are independent. In this case tunneling in the $x$ direction
occurs according to the WKB one-dimensional scenario.

The eigenstate with the underbarrier energy $\gamma-1$ in the total potential $V_{0}(x)+u(x,y)$ can be considered as a superposition of 
eigenstates of the potential $V_{0}(x)$: (i) the underbarrier WKB type state and (ii) propagating waves with overbarrier energies. In the 
first order of a perturbation theory with respect to $u(x,y)$ the processes (a) in Fig.~\ref{fig12} participate in formation of the exact
underbarrier state. Subsequent scattering processes, of next orders of the perturbation theory, modify the propagating waves as shown in 
Fig.~\ref{fig12} by the paths (b). The term ``eigenstate'' is used since one can ignore an exponentially small leakage across the barrier. 

A principal question is that how strong is a contribution of the propagating waves to the total wave function. First of all, if $u(x,y)$ is 
not small, each partial propagating wave is also not small and does not decay exponentially with distance. But there is an opposite 
tendency. When the waves are distributed smoothly in energy and directions the mutual interference reduces their contribution to the wave 
function down to an exponentially small value.

However, a completely different scenario can be realized when a tangent momentum in the prebarrier region is not zero as in 
Fig.~\ref{fig1}(b). In our case this corresponds to a finite momentum $p_{y}$ in the wire ($x=0$). The finite tangent momentum may result in
a new phenomenon in the distribution of propagating waves. A classical trajectory in the potential $(-x)$ (at a not small positive $x$ where
the impurity potential is weak) can be generally written as
\begin{equation} 
\label{110}
x=p^{2}_{y}+1-\gamma+\frac{(y-ib)^{2}}{4p^{2}_{y}}.
\end{equation} 
Here $ib$ is a reflection point in $y$, $p_{y}$ is a tangent momentum, and the underbarrier energy $\gamma-1$ is negative. The trajectory 
(\ref{206}) is a particular case of the form (\ref{110}). 

A type of the trajectory (\ref{110}) strongly depends on $p_{y}$. At zero tangent momentum the trajectory (\ref{110}) is degenerated into the
line $y=0$ when $b=0$. This is the main underbarrier path of the same type as in Fig.~\ref{fig1}(a). At a finite $p_{y}$ the curves 
(\ref{110}) become two-dimensional. They are analogous to rays in geometrical optics. When the optical rays are not parallel they are 
reflected by certain curves which are caustics \cite{LANDAU1}. The same caustic phenomenon in the space $\{x,{\rm Im}y\}$ occurs with the 
trajectories (\ref{110}) as follows from the results of Secs.~\ref{sol} and \ref{2D}. 

From the standpoint of the three-dimensional space $\{x,{\rm Re}y,{\rm Im}y\}$, the caustic curve pierces the physical two-dimensional plane 
at the point $\{x_{0},\Delta\}$ as shown in Fig.~\ref{fig6}. In a vicinity of this point a distribution of the propagating waves becomes not 
smooth (as close to a caustic in optics) and their interference cannot now lead to a mutual compensation in contrast to a smooth 
distribution. As a result, a significant fraction of the prebarrier density is carried away by propagating waves. Therefore, a barrier 
penetration is determined in that case not by a conventional underbarrier mechanism, as in Fig.~\ref{fig1}(a), but by interference of the 
propagating waves, as in Fig.~\ref{fig1}(b)). 

We outline by general arguments the nature of undebarrier interference. To draw more exact conclusions we use the rigorous method of 
summation of various propagating waves in Fig.~\ref{fig12}. This method is generic with a saddle formalism which allows to collect rapidly 
oscillating functions. In our case this is the semiclassical approach based on the equation of Hamilton-Jacobi used in Sec.~\ref{ham-jac}. 
\section{TWO ENTANGLED PARTICLES}
\label{entangl}
A process of tunneling in two dimensions is sketched in Fig.~\ref{fig1}. This tunneling scenario can also be realized in a two-dimensional 
potential barrier of the type
\begin{equation} 
\label{26}
V(x,y)=\frac{V}{\cosh^{2}(x/a)}+\frac{V}{\cosh^{2}(y/a)},
\end{equation}
when the particle energy $E$ is less than $V$. Suppose tunneling to occur symmetrically, that is in the direction $x=y$ in the plane 
$\{x,y\}$. A tangent momentum is in the direction $x=-y$, which is perpendicular to tunneling. One can show (we will present calculation 
elsewhere) that in this case a phenomenon of underbarrier interference also takes place. The phenomenon is generic with one investigated 
above and resulting in the enhanced tunneling. 

The problem of a particle in the two-dimensional potential (\ref{26}) can be considered from another standpoint. Namely, there are two 
different particles described by the coordinates $x$ and $y$ and  moving in identical one-dimensional potentials. In classical mechanics the 
two particle are completely independent and can be spatially separated by an arbitrary long distance. 

In quantum mechanics everything depends on a type of the wave function. The wave function of two particles 
$\psi(x,y)=\psi_{1n}(x)\psi_{2n}(y)$ describes independent motions through each separate barrier. Suppose now that the two particles are in 
the entangled state 
\begin{equation} 
\label{27}
\psi(x,y)=\sum_{E_{1n}+E_{2n}=E} c_{n}\psi_{1n}(x)\psi_{2n}(y)
\end{equation}
related to a momentum of the total system in the direction $x=-y$ and to some classical turning point along the line $x=y$. The wave 
function (\ref{27}) is analogous to one for a single particle in the two-dimensional potential (\ref{26}) with the total energy $E$. 
Therefore, the two particles can penetrate the barrier instantaneously and with a not small probability despite they are separated by an 
arbitrary distance. So by means of the second particle the conventional WKB mechanism can go over into one with the underbarrier 
interference.  

This scheme provides some sort of quantum communication when one can induce a penetration through a barrier of the ``controlled'' 
particle which is distant from the ``controlling'' one. A role of particles with barriers can be played by two Josephson junctions 
\cite{BARONE}. With no interference effects tunneling in Josephson junctions was initially observed in Ref.~\cite{CLARKE}. We do not discuss
here a way to create the properly entangled state of two junctions. 
\section{DISCUSSIONS}
The counterintuitive phenomenon of easy penetration through barriers which are nontransparent, according to WKB theory, is proposed. A 
magnetic field is zero, the potential barrier is static and non-homogeneous in the direction perpendicular to tunneling. 

When the applied electric field is less than the certain value ${\cal E}_{R}$, decay of the prebarrier state is exponentially weak and, in 
principle, similar to a conventional WKB decay. 

At a larger field, ${\cal E}_{R}<{\cal E}$, a scenario of barrier penetration becomes different. An initially created prebarrier state will 
be transferred fast (during the time of an inverse barrier height) outside the barrier. 

An essential point of the phenomenon is a generation of propagating waves under the barrier when, prior to tunneling, a particle has a 
momentum perpendicular to a tunneling direction (a tangent momentum). Tunneling from the ground state is of a conventional WKB nature since 
a tangent momentum is zero in this case. In presence of a finite tangent momentum the propagating waves do not cancel each other completely 
by interference and the surviving part provides a not small output from under the barrier. 

The state outside the barrier, described in Sec.~\ref{sol}, is a static packet. The top of the packet follows the classical trajectory 
(\ref{103d}) where it keeps a constant amplitude. If to account for quantum effects, beyond the  Hamilton-Jacobi approach, its amplitude 
reduces at a larger $|y|$. The momentum, associated with the packet, is gained from the quantum wire. The momentum is directed towards the 
barrier at $y<0$ and it is opposite at $0<y$. The packet is generated at the points $\{x_{0},\pm\Delta\}$ where the caustics pierce the 
physical plane $\{x,y\}$, Fig.~\ref{fig4}(b). At $\Delta<|y|$ the packet outside the barrier and the initial branch 1-1 are disconnected. At
$|y|<\Delta$ the initial branch softly undergoes into the packet, the curve 1-3 in Fig.~\ref{fig4}(b).

An exact analytical solution is not obtained. However, on the basis of the Stokes theory it is proved that the semiclassical exponent is a
main part of the exact wave function in the problem. A prefactor does not unexpectedly jump and, therefore, it plays a secondary role.   

Suppose the underbarrier state to be artificially cut off at the initial moment so that there is no the outer packet. After a release the 
dynamical state is developed, when transitions in the entire spectrum occur. The typical energy scale involved is the barrier height 
$u_{0}$ and therefore the time scale of restoring of the branch 1-3 is of the order of $\hbar/u_{0}$. This is a short (nonsemiclassical) 
time. Under increase of the electric field, above the certain threshold ${\cal E}_{R}$, the prebarrier region gets emptied fast (instant 
tunneling). ${\cal E}_{R}$ depends on energy of a decaying state in the quantum wire. For a higher energy the threshold ${\cal E}_{R}$ is 
less. 
                                      
As we see in Sec.~\ref{imp}, the scattering of underbarrier waves by the impurity also results in the wave packet propagating outside the 
barrier. The both situations, described in Secs.~\ref{sol} and \ref{imp}, are generic and can be unified as scattering of underbarrier waves
by non-homogeneities. The phenomenon of underbarrier interference provides a different aspect in study of tunneling through barriers with
impurities \cite{LIF,SHKL1,MESH,SHKL2} since an individual impurity may strongly increase the tunneling rate.

Tunneling through a barrier with an impurity can be unusual even for zero tangent momentum of a particle at a prebarrier region. But in 
this case the impurity should be dynamic. A typical example is alpha decay of a nucleus assisted by an incident proton \cite{IVLEV3}. The 
moving proton plays a role of a dynamic impurity. Caustics are formed in the joint space of alpha particle and proton coordinates. Due to 
interference in the $\{\alpha, p\}$ system, moving protons can increase the alpha decay rate making it not exponentially small. 

Tunneling across a one-dimensional barrier with a nonstationary slope also can be strongly enhanced \cite{IVLEV4}. Quanta absorption and
emission result in an underbarrier phase analogous to one produced by the prebarrier momentum $p_{y}$ in two dimensions. This leads to a 
phenomenon, analogous to caustics in Fig.~\ref{fig6}, which prevents a mutual cancellation of the propagating states.

The underbarrier interference is also a feature of tunneling in multi-dimensional systems. An elastic string in a washboard potential can 
tunnel from some valley (the initial valley) to another one (the basic valley) through a potential barrier 
\cite{VOL,STONE,COLEMAN,COLEMAN1,COLEMAN2,MELN}. If the string has a momentum along the initial valley and the string or the washboard is 
not homogeneous in that direction, an enhancement of tunneling, caused by the interference, may occur. Calculations will be published 
elsewhere. As a result of tunneling, the certain part of the string appears in the basic valley and extends along its direction. In presence
of a friction the part, coming to the basic valley, extends with a constant velocity. From the standpoint of the basic valley, this 
phenomenon looks as a violation of energy conservation if the basic valley is thought to be a single one in the general potential.

The enhanced penetration through a classical barrier, caused by the underbarrier interference, can be observed in various natural and 
artificial tunneling systems. Below some of them are mentioned. (1) Tunneling from a wire or a film. Their non-homogeneities 
should satisfy the certain not very rigorous conditions of the type described in Sec.~\ref{sol}. (2) Tunneling from a homogeneous wire or 
film through a barrier with impurities. (3) A system of two coupled Josephson junctions can manifest a not weak tunneling across a 
non-transparent barrier as a result of the underbarrier interference. (4) Tunneling in two distant systems which are properly entangled 
(Sec.~\ref{entangl}). (5) Ionization of a molecule in an electric field can exhibit the phenomenon of interference if the molecule rotates 
around an axis perpendicular to the electric field. (6) The proton assisted alpha decay of nuclei is also an example of an experimental 
investigation.
               
We studied above a pure Hamiltonian system. An interference in this case differs from one in presence of friction, for example, phonons 
\cite{LEGGETT,WEISS}. A role of friction is worth to be studied. 
\section{CONCLUSIONS}
Quantum tunneling through a two-dimensional static barrier becomes unusual when a momentum of an electron has a tangent component with 
respect to a border of the prebarrier region. If the barrier is not flat a fraction of the electron state is waves propagating away from the 
barrier. When the tangent momentum is zero a mutual interference of the waves results in an exponentially small outgoing flux. The finite 
tangent momentum destroys the interference due to formation of caustics by the waves. As a result, a significant fraction of the prebarrier 
density is carried away from the barrier providing a not exponentially small penetration even through an almost classical barrier. The total
electron energy is well below the barrier. 
\acknowledgments 
I thank A. Barone, G. P. Berman, G. Blatter, M. I. Dykman, V. B. Geshkenbein, V. Gudkov, S. A. Gurvitz, I. A. Larkin, G. Pepe, and 
A. V. Ustinov for valuable discussions of the paper and related topics.

\end{document}